\documentclass[twocolumn,showpacs,aps,prb]{revtex4}
\usepackage{graphics}
\usepackage{epsfig}
\DeclareTextSymbol{\degre}{OT1}{23}
\begin{document}
\title{Surface state bi-polarons formation on a triangular lattice in the sp-type alkali/Si(111) Mott insulator}
\author{L.A. Cardenas$^1$, Y. Fagot-Revurat$^1$, L. Moreau$^1$, B. Kierren$^1$, and D. Malterre$^1$}
\address{$^1$Institut Jean Lamour, UMR 7198,
Nancy Universit\'e/CNRS, B.P. 239 F-54506 Vand\oe
uvre-l\`es-Nancy, France}
\date{\today}
\begin{abstract}
We report on new LEED, STM and ARPES studies of alkali/Si(111)
previously established as having a Mott insulating ground state at
surface. The observation of a strong temperature dependent
Franck-Condon broadening of the surface band together with the
novel $\sqrt{3}\times\sqrt{3}\rightarrow2(\sqrt{3}\times\sqrt{3})$
charge and lattice ordering below 270 K evidence a surface charge
density wave (SCDW) in the strong e-ph coupling limit
($g\approx8$). Both the adiabatic ratio
$\hbar\omega_0/t\approx0.8$ and the effective pairing energy
$V_{eff}=U-2g\hbar\omega_0\approx-800$ $meV$ are consistent with
the possible formation of a bi-polaronic insulating phase
consisting of alternating doubly-occupied/unoccupied dangling
bonds as expected in the Holstein-Hubbard model.

\end{abstract}

\maketitle

%
The question of the relevance of electron-phonon coupling (EPC) in
the physical properties of strongly correlated systems has been
recently addressed for various materials like
manganites\cite{Millis98}, alkali-doped
$C_{60}$\cite{Gunnarson97}, vanadium oxides\cite{Limelette03} and
especially cuprates for which the interplay between charge, spin
and lattice degrees of freedom is still unsolved\cite{Gweon04}.
From the theoretical point of view, the half-filled
Holstein-Hubbard model, including both electron-electron and
electron-phonon coupling, has been intensively studied
\cite{Meyer02,Berger95,Fehske04,Jeon04,Sangiovanni05}. In the
weak-coupling limit, a correlated/polaronic metal takes place. In
the strong-coupling limit, the competition between electronic and
lattice effects leads to an effective interaction
$V_{eff}=U-2g\hbar\omega_{0}$ whose sign determines the ground
state : a Mott-Hubbard insulating one (MHI) takes place at large U
whereas an attractive e-e interaction can occur for large g
resulting in a quantum phase transition from a MHI to a
bi-polaronic insulating phase (BPI). Therefore, localized electron
pairs are trapped by their self-induced lattice deformation
leading to the formation of a charge density wave state implying a
doubling of the lattice parameters\cite{Gobel94} as early observed
in 3D titanium oxides\cite{Lakkis76}. Lowering the dimension,
itinerant small bipolarons on triangular lattices have been
recently predicted to form a Bose-Einstein condensate giving
possibly rise to high temperature superconductivity\cite{Hague07}.

Peculiar Mott insulators implying sp states can be found at
semi-conducting surfaces. Indeed, the reduction of the bandwidth W
and of screening effects at surface have been first proposed to
explain the Mott transition leading to the insulating properties
of K/Si:B\cite{Weitering97,Hellberg99}. Then, such a Mott state
has been identified as a generic property of semiconducting
surfaces with dangling bonds (DB) organized in a $\sqrt{3}-$
surface reconstruction i.e., presenting a triangular
pattern\cite{Flores99}, such as SiC(0001)\cite{Santoro99} and
C/Si(111)\cite{Profeta05}. Recently, surface science tools have
been used to solve the metal/insulator transition
in Sn/Ge(111)\cite{Cortes06,Colonna08} and
Sn/Si(111)\cite{Modesti07}. This triangular topology has been
predicted to induce exotic magnetic
phases\cite{Weitering97,Santoro99,Trumper04} and even
superconductivity upon doping\cite{Profeta07}. Otherwise, weak EPC
theories have been proposed to explain several surface
reconstructions as well as ARPES/STS spectral
features\cite{Gonzalez06,Barke06,Berthe06}. Weitering and
co-workers have pointed out the possible role of EPC in the case
of K/Si:B\cite{Weitering97}. Angle-resolved photoemission
spectroscopy (ARPES), low-energy electron diffraction (LEED) and
scanning tunnelling microscopy (STM) have been combined to solve
the fundamental nature of the K/Si:B interface. Indeed, the novel
$\sqrt{3}\times\sqrt{3}\rightarrow2(\sqrt{3}\times\sqrt{3})$
insulating-insulating phase transition observed here together with
the evidence of a phonon-dressed spectral function allow us to
identify the ground state of K/Si:B as a bi-polaronic insulator
rather than a Mott-Hubbard one. Our results imply that K/Si:B is
to our knowledge the first experimental realization of a BPI phase
on a triangular lattice showing that $in$ $fine$ the EPC drives
the electronic properties in this correlated material.

%
%
Boron-enriched Si(111) substrates ($\rho\approx10^{-3}$
$\Omega\times cm$) have been shortly annealed at $1450$ $K$
followed by a few hours heating at $1000$ $K$ to favor the
segregation of B-atoms in the pentavalent $S_{5}$ sub-surface
site\cite{Baumgartel99}. A nearly free of defects
$\sqrt{3}\times\sqrt{3}R30$ surface reconstruction ($a^{'}=6.66$
$\AA$) with less than 0.015 B vacancies per nm$^2$ has been
usually obtained as checked by STM.
\begin{figure}[t]
\begin{center}
\scalebox{0.45}{\includegraphics{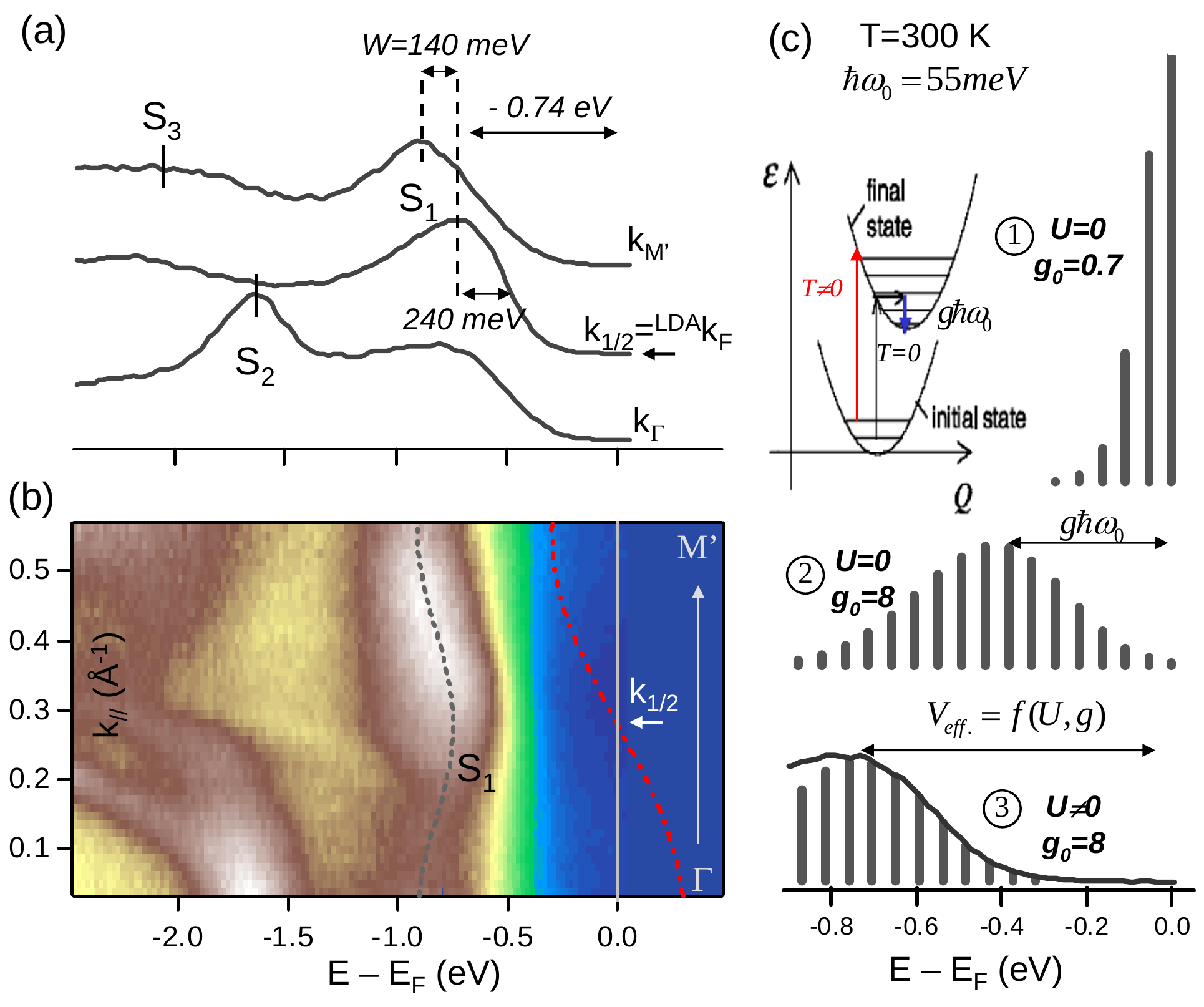}}
\end{center}
\vspace{-0.7cm}
\caption{\label{one} EDC's (a) and ARPES intensity
map (b) measured along $\Gamma-M'$ in 0.33 ML K/Si:B at 300 K; red
and black lines materialize respectively the tight-binding and the
experimental dispersion curves; (c) Franck-Condon spectral
function and adjustment of the experimental leading edge of the
$k=k_{1/2}$ spectra (continuous black line)}
\end{figure}
On the Si(111) surface, the dangling bonds are reorganized in a
$7\times7$ surface reconstruction with a weakly dispersive ad-atom
half-filled surface state ensuring its slight metallic
character\cite{Barke06}. At the opposite, the B-enriched Si(111)
substrate exhibits a semiconducting behavior, DB electrons are
trapped by boron impurities leaving an empty surface
state\cite{Grehk92}. The B-enriched Si(111) band structure has
been shown to be well described by ab initio
calculations\cite{Shi02}. The K/Si:B interface was prepared at 300
K by evaporation of K atoms from a SAES getter with a deposition
rate of 0.13 monolayer/mn in a vacuum better than
$5\times10^{-10}$ $mbar$. The room-temperature (RT) saturation
coverage, defined as one K atom per reconstructed unit cell (1/3
of alkali monolayer) has been identified by in-situ monitoring of
Auger Si, B and K lines. According to previous studies, the
room-temperature sticking coefficient falls down after the
completion of the first layer i.e. at the saturation coverage of
0.33 ML K/Si:B\cite{Weitering93}. The LEED pattern obtained at
saturation indicates a well-defined $\sqrt{3}\times\sqrt{3}R30$
surface reconstruction in agreement with the K chemisorption in
the highly coordinated $H_3$ site\cite{Shi02}. Previous studies of
core-levels have established a charge transfer occurring from the
$K4s$ orbital to the silicon ad-atom at surface, leading to a new
DB state induced by K adsorption\cite{Ma90}. Naively, with one
electron per DB-orbitals, a half-filled metallic state was
expected to be formed on a triangular lattice as shown by
ab-initio calculations\cite{Shi02}.

ARPES intensity map and energy dispersion curves (EDC) obtained at
the saturation coverage with an energy and momentum resolution of
$10$ $meV$ and $0.04\AA^{-1}$ are displayed in figure \ref{one}.
According to previous results\cite{Weitering93}, a well-defined
K-induced surface state labelled $S_1$ is developed in the
semiconducting gap, its spectral weight being related to the
amount of K deposited. The $S_2$ and $S_3$ bands have been
ascribed to rest-atom and backbond surface states\cite{Shi02}. As
a new insight, the surface state dispersion has been accurately
measured along the $\Gamma-M^\prime$ ($\Gamma-K^\prime$) high
symmetry direction of the $\sqrt{3}\times\sqrt{3}$-hexagonal SBZ.
In qualitative agreement with calculations, the surface band
dispersion presents minima at $M^\prime$ ($k=0.55$ $\AA^{-1}$,
$E-E_F=-880$ $meV$) and $K^\prime$ ($k=0.63$ $\AA^{-1}$,
$E-E_F=-900$ $meV$) with a positive effective mass. Nevertheless,
contrary to the tight-binding predictions (dashed-dotted line in
figure \ref{one}-b) or ab initio calculations\cite{Shi02}), a
clear band-folding occurs close to half of the BZ
($k=k_{1/2}\approx0.27$ $\AA^{-1}\approx k_{F}^{LDA}$) leading to
a second minimum in the band dispersion at the zone center
($\Gamma$) and the opening of a gap larger than $500$ $meV$. A
similar behavior has been observed along $\Gamma-K'$ leading to
the reduction by a factor of four of the RT apparent surface BZ.
However, the spectral weight is mainly located in the unfolded
part of the one-electron surface band ($k>k_{1/2}$). In addition,
a strongly renormalized experimental occupied bandwidth
$W_{exp}^{occ.}=140$ $meV$ have been deduced compare to the ab
initio value of $300$ $meV$ expected for the occupied
part\cite{Shi02}. This large gap has been first interpreted as
resulting from a correlation induced metal/insulator
transition\cite{Weitering97}. Indeed, the Harrison criterion
$U/W>>1$ was supposed to be full-filled at surface for DB
electrons with $U\approx1.5$ $eV$\cite{Weitering97} and
$W_{LDA}^{full}\approx0.6$ $eV$ ($t\approx0.07$ $eV$)\cite{Shi02}
giving rise to a surface-induced Mott transition\cite{Hellberg99}.
The binding energy $E_B^{S1}=0.75$ $eV\approx U/2$
agrees well with the Mott insulator model. K/Si:B was claimed to
be the first 2D-Mott-Hubbard material based on sp-band, the
triangular topology being speculated to give rise to singular
magnetic properties\cite{Weitering97,Trumper04}.

The novel apparent band folding at $k_{F}^{LDA}\approx k_{1/2}$
evidenced here does not contradict the MI model\cite{Fehske04}.
Nevertheless, one should remark on figure \ref{one}-a the strong
intrinsic broadening of the SS at 300 K even considerably larger
than the bandwidth. Weitering and co-workers have initially
proposed to interpret this broadening as a possible Franck-Condon
(FC) envelop due to a strong EPC but did not prove
it\cite{Weitering97}. For localized electrons, the FC model
describes the coupling of an electron to a single harmonic
oscillator\cite{Mahan81} as depicted in the figure \ref{one}-c for
the specific case of K:SiB at 300 K. As a general feature of EPC,
the PES spectra contain phonon side-bands at
$E_k=E_k^0-n\hbar\omega_0$ (n=0,1,...) reflecting the screening of
the photo-hole by localized vibrational excitations\cite{Mahan81}.
The T=0 intensity of the $n^{th}$ transition is therefore given by
the FC factor $I(n)=g^{n}\exp(-g)/n!$, $g$ characterizing the
strength of the EPC. In the weak coupling limit $g<1$, the
zero-phonon transition, at the free binding energy $E_k=E_k^{0}$,
is the most pronounced, other transition causes a broadening of
the PES line (case 1, figure \ref{one}-c). In the strong coupling
limit $g>1$, the spectral weight is redistributed over a wide
energy range with a maximum at $E_k=E_k^{0}-g\hbar\omega_0$ (case
2, figure\ref{one}-c). Additional extrinsic broadening usually
prevents the clear observation of multi-phonon
side-bands\cite{Berthe06}. Therefore, for $g>>1$, a
nearly-Gaussian incoherent broad spectral intensity is
experimentally observed whose width is determined by the average
number of phonons involved in the final state at T=0. Moreover, at
finite temperature ($k_{B}T\geq\hbar\omega_0$), the lineshape is
broadened due to the contribution of excited vibrationnal
levels\cite{Mahan81}. This has been commonly used as a fingerprint
for the EPC signature even in strongly correlated
materials\cite{Shen07}.
\begin{figure}[b]
\vspace{-0.5cm}
\begin{center}
\scalebox{0.6}{\includegraphics{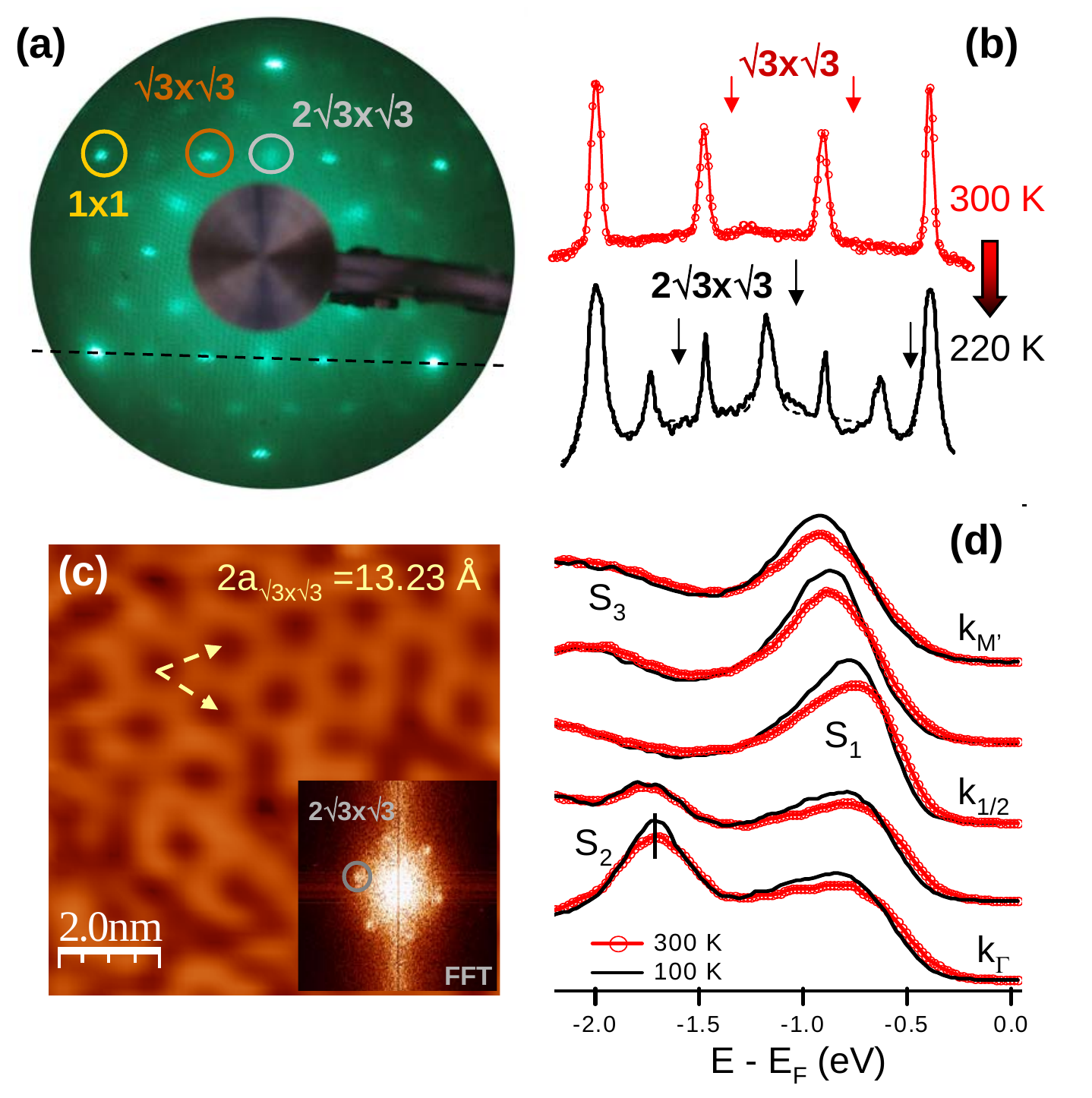}}
\end{center}
\vspace{-0.5cm} \caption{\label{two} (a) LEED pattern obtained at
220 K, (b) profiles obtained at 300 and 220 K; (c) STM image
(V=1.3 V, I=0.2 nA) obtained at 80 K; inset : FFT obtained on a
80nmx80 nm image; (d) ARPES spectra obtained at 300 K and 100 K}
\end{figure}
Assuming the FC broadening as the dominant contribution to the
experimental spectral width, an attempt has been made to adjust
HWHM of the low energy tail of the room-temperature $k=k_{1/2}$
EDC (case 3, figure\ref{one}-c). A large coupling constant
$g\approx8$ and a phonon energy $\hbar\omega_0=55$ $meV$ have been
deduced consistent with those obtained on similar semiconducting
surfaces\cite{Barke06,Berthe06}. Such a strong EPC might possibly drive
surface reconstructions as presented in the following.

\begin{figure}[t]
\begin{center}
\scalebox{0.4}{\includegraphics{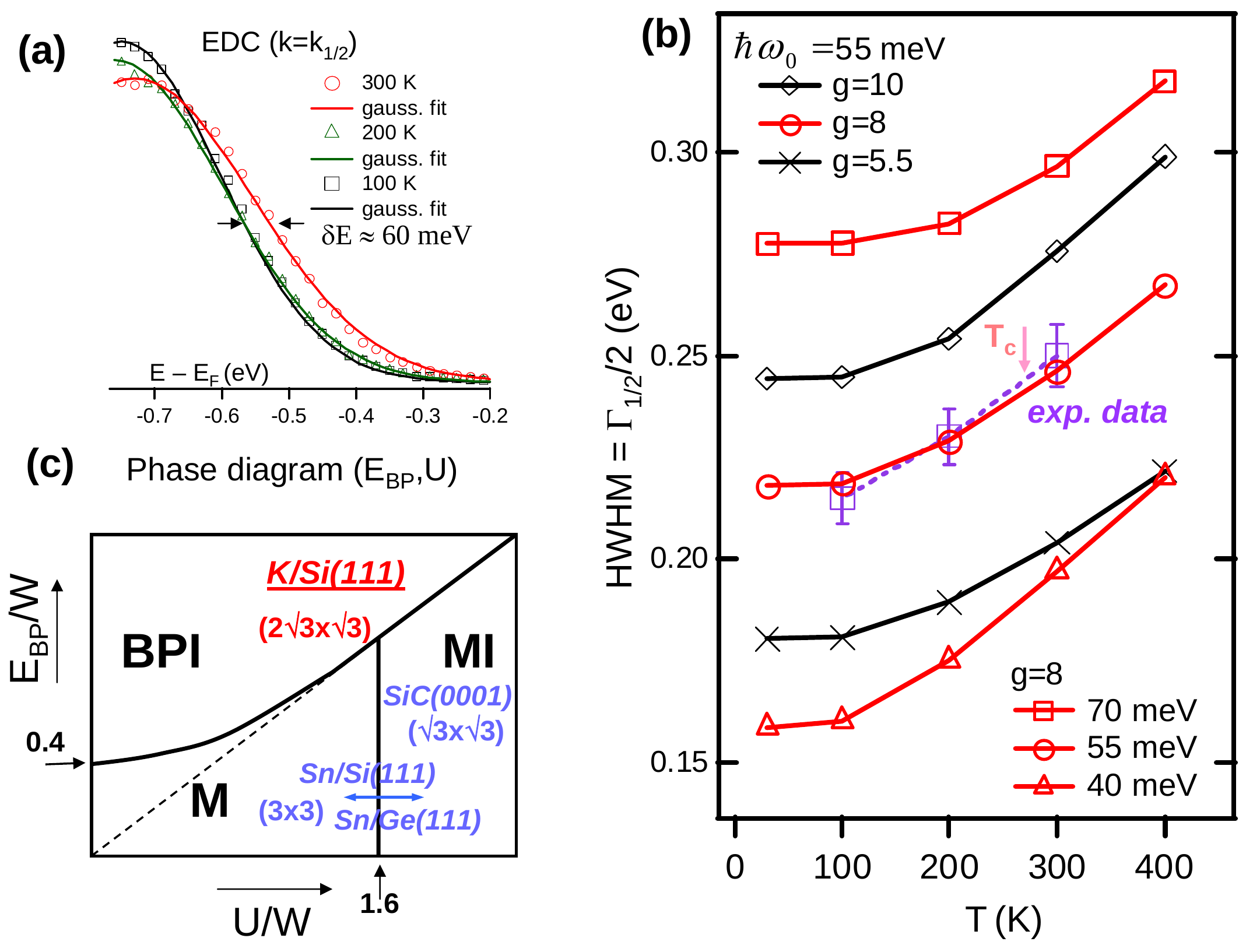}}
\end{center}
\vspace{-0.5cm} \caption{\label{three} EDC's(k=k$_{1/2}$(a),
calculated FC broadening compared to the experimental gaussian
fits (b); general (g,U) phase diagram in the Holstein-Hubbard
model\cite{Jeon04} (c)}
\end{figure}

On one hand, the LEED pattern obtained at 220 K displayed in
figure \ref{two}-a exhibits a novel $2(\sqrt{3}\times\sqrt{3})$
phase at surface. Corresponding profiles obtained at 300 K and 220
K presented in figure \ref{two}-b indicate a clear phase
transition involving a lattice distortion i.e. a doubling of the
hexagonal lattice parameters. The narrow 220 K-diffraction peaks
indicate that the new structural ordering corresponds to a well
established long range ordered phase. The transition has been
identified around $T_{c}\approx270$ $K$. However, the long range
ordering, the transition temperature, and a moderate hysteretic
behavior depend on the amount of Boron vacancies at surface acting
as defects. A more detailed study will be presented elsewhere. In
addition to the lattice distortion, a clear charge density
modulation is evidenced by our STM study of the low-T phase
presented in figure \ref{two}-c. Despite a small disorder, the
2D-FFT spectrum obtained on a 80x80 nm$^2$ STM image exhibits a
well-defined $2(\sqrt{3}\times\sqrt{3})$ peaks evidencing the
quadrupling of the unit cell at surface. On the other hand,
quantitative modifications of the surface band have been observed
as illustrated in figure \ref{two}-d. A systematic redistribution
of spectral weight to higher binding energies is clearly evidenced
including a $\delta E\approx60$ $meV$ shift of the SS leading edge
between 300 K and 200 K, this one remaining constant below 200 K
(see also in figure \ref{three}-a). As evidenced in the figure
\ref{two}-d, other bands remain unaffected down to 100 K
indicating there are not charge or photovoltage effects and
ensuring that only the K-induced surface state is involved in this
surface transition. The bandwidth is slightly reduced together
with a stabilization of the electronic surface band in agreement
with a more insulating LT-phase. Moreover, we would like to point
out the clear narrowing of the intrinsic width of the surface band
over a wide temperature range, even far from the transition.
Indeed, following the procedure of Shen and
co-workers\cite{Shen07}, a gaussian adjustment of the HWHM (the
low energy tail) of the k=k$_{1/2}$ ARPES spectra have been made
for 100, 200 and 300 K as presented in figure\ref{three}-a. The
result has been displayed as function of the temperature in figure
\ref{three}-b and compared with those we have calculated from the
FC model varying g, $\hbar\omega_0$ and taking into account the
temperature. Again, a good agreement between experimental data and
calculations is achieved for $g\approx8$ and
$\hbar\omega_0\approx55$ $meV$. Therefore, the novel set of data
presented in this letter strongly supports the existence of a
$\sqrt{3}\times\sqrt{3}\rightarrow2(\sqrt{3}\times\sqrt{3})$
insulating to insulating surface phase transition involving a
lattice dimerization, a charge modulation and a net energy gain
for surface state electrons. Contrary to previous studies, they
shed on light the strong relevance of the electron-phonon coupling
to understand the K/Si:B ground state.

The absence of Fermi surface, the large gap measured in both
phases and the reconstructed LT-phase contradict previous ab
initio predictions\cite{Shi02}. Therefore, g and U are supposed to
possibly play a role and the data have been analyzed in the
framework of the Hostein-Hubbard model at half-filling whose
generic T=0 phase diagram is depicted in figure \ref{three}-c
(taken from DMFT calculations\cite{Jeon04}). On one hand, a Mott
state is achieved for $U/W\geq1.6$ but does not imply a lattice
symmetry breaking. However, it can be favored by a lattice
reconstruction ($3\times3\rightarrow\sqrt{3}\times\sqrt{3}$ in
Sn/Ge(111)\cite{Cortes06}) or by a precursive CDW transition as
observed for $1T-TaSe_2$\cite{Perfetti03}. On the other hand, in
the Holstein model(U=0), a soft phonon induces a lattice
instability beyond a critical e-ph coupling leading to the
formation of a CDW state involving an increase in the unit cell in
qualitative agreement with our experimental
data\cite{Meyer02,Fehske04,Jeon04}. A high g and an adiabatic
ratio close to 1 suggest the strong e-ph coupling limit is reached
here leading to the formation of a bi-polaronic insulator (BPI).
Indeed, the pairing energy defined by $E_{BP}=2g\hbar\omega_{0}$
full-filled the criterion $E_{BP}/W=1.2>0.4$ for establishing the
BPI phase\cite{Jeon04}. Assuming the 4s alkali electron is mainly
transferred to the DB orbital (Si ad-atom)\cite{Weitering93}, the
quadrupling of the unit cell is easily obtained by alternating
doubly- and un-occupied DB-site as expected for the BPI
state\cite{Fehske04,Gobel94}. As for BCS theory of
superconductivity, the full gap corresponds to twice the pairing
energy for bounded two electrons\cite{Meyer02}. Therefore, the
occupied part of the gap is theoretically expected to be
$E_{BP}\approx880$ $meV$\cite{Meyer02}. However, the effective
pairing energy might be reduced by the coulombic repulsion $U$
leading to an occupied gap given by $2g\hbar\omega_{0}-U/2$.
Hence, a weak $U\approx160$ $meV$ might explained our
LT-experimental (occupied)gap of less than $800$ $meV$. The BPI
phase is predicted to present a poorly dispersive back-folded
phonon-dressed ARPES spectral function\cite{Gobel94,Fehske04} in
agreement with the results presented in this letter. Therefore, as
suggested in the phase diagram presented in figure \ref{three}-c,
the BPI ground state might be reached in 0.33 ML K/Si(111)-B
contrary to other $\sqrt{3}$-semiconducting surfaces. Such a BPI
phase has been also proposed to describe the ground state of 0.25
ML Na/GaAs(110) on the basis of DFT
calculations\cite{Pankratov93}. This suggests the BPI ground state
as a common property of ultra-thin alkali layer deposited on
semiconducting substrates. Unlike weak-coupling theories, Fermi
surface nesting properties are not necessary to reach the long
range ordered state in the strong coupling limit\cite{Gobel94}.
Hence, the high temperature phase might not be necessary metallic.
Indeed, a precursive short range ordered intermediate insulating
phase, defined as a bi-polaronic liquid, has been observed a long
time ago in $Ti_4O_7$\cite{Lakkis76}, the long range
$Ti^{3+}-Ti^{4+}$ charge ordering occurring only below 130 K.
The precursive diffusive background observed in LEED patterns
together with the well established surface state band-folding
above $T_c$, could be indications for a short range ordered BPI
phase at 300 K, the lifetime characterizing polaronic excitations
being short enough to experiment the fluctuating surface
reconstruction. This is an open question. Due to their bounded
nature, bi-polarons should carry a singlet spin state making these
surfaces non-magnetic contrary to previous
speculations\cite{Weitering97}.

Substitution with other alkali should provide a
way to explore the (g,U) phase diagram
of the 2D-Holstein-Hubbard model on a triangular lattice. Many
interesting aspects such as isotopic effects, core level signature of the charge
ordering, direct measurements of the typical phonon energy by EELS should be
investigated in the future. Finally, theoretical and experimental works
carried on these SC surfaces these last ten years have
evidenced their extreme proximity with the
Mott-Hubbard physics in presence of a coupling to the lattice\cite{Profeta07}. Doping these half-filled systems could
provide the opportunity to investigate novel ground state at
surface including high Tc superconductivity.

We gratefully acknowledge to A.E. Trumper, L.O. Manuel, F. Flores,
J. Ortega and C. Tournier for our stimulating discussions.

\end{document}